# Liquid actuated gravity experiments

Massimo Bassan[1,2], Martina De Laurentis[3,4], Rosario De Rosa[3,4], Luciano Di Fiore[4*], Luciano Errico[3,4], Fabio Garufi[3,4], Aniello Grado[4,5], Yury Minenkov[1], Giuseppe Pucacco[1,2] and Massimo Visco[2,6]

[1] Dipartimento di Fisica, Università di Roma Tor Vergata, I-00133 Roma, Italy

[2] INFN - Sezione Roma2, I-00133 Roma, Italy

[3] Dipartimento di Fisica, Università di Napoli "Federico II", I-80126, Napoli, Italy

[4] INFN - Sezione di Napoli, I-80126, Napoli, Italy

[5] INAF - Osservatorio Astronomico di Capodimonte, I-80131 Napoli, Italy

[6] INAF - Istituto di Astrofisica e Planetologia Spaziali, I-00133 Roma, Italy

E-mail: luciano.difiore@na.infn.it

**Abstract**

We describe a new actuation technique for gravity experiments based on a liquid field mass. The Characterizing idea is to modulate the gravity force acting on a test mass by controlling the level of a liquid in a suitable container. This allows to obtain a periodical gravity force without moving parts (except the liquid level) close to the TM. We describe in detail the most relevant aspects of the liquid actuator and discuss how it can be used in gravity experiments. In particular we analyse an application to test the inverse square law in the mm to cm distance region.

---

[*] Corresponding author

# 1. Introduction

Gravity experiments are generally based on one (or more) field mass (FM) interacting with a test mass (TM), frequently suspended to a torsion pendulum. The measured physical effect can be the change in either pendulum orientation or period of swing. We can mention as examples experiments to measure the gravitational constant *G* and to check possible violation of the inverse square law (ISL) and of the principle of equivalence. A review of these experiments can be found in references [1][2][3]. Many possible geometries and materials have been used for both FM and TM in various experiments, where generally the relative TM-FM distance is changed from a near to a far position. In particular, a few years ago, liquid FMs made of mercury in cylindrical containers have been used for an experiment to measure *G* [4], by moving them between two different positions with respect to two TMs suspended to a beam balance. Modulation of the applied force is a standard technique often used in gravity experiments in order to improve S/N ratio by coherent detection. This is generally achieved by changing the position of a FM with respect to the TM [4][17][6][7].

In this paper, we describe a Liquid Actuated Gravity (LAG) apparatus, where the gravitational force is modulated without moving parts close to the apparatus, but instead by changing, in a controlled and repeatable way, the level of a liquid used as FM.

The advantage of using a liquid FM, with respect to solid one is twofold.

First of all, the density of, a pure, stationary liquid is generally much more uniform than that of a solid. The main density excursion is due to isobaric pressure gradient and liquid compressibility. For an Hg FM with height of 10 cm the density variation is $\sim 4 \cdot 10^{-7}$ to compare to typically $\sim 10^{-4}$ for inhomogeneity in solids.

A second important advantage is that we can modulate the gravitational force from the maximum (upper liquid level) to the minimum (lower liquid level) while keeping unchanged all the experimental set-up close to the TM. This allows obtaining a large signal without changing the FM position. With a solid FM with comparable size and volume, to get the same signal amplitude one should move periodically the FM by a large amount compared to FM size and FM-TM distance. In this case, all the moving parts of the mechanism would contribute to modulate the force making it more difficult both the evaluation of the expected signals and its comparison with experimental data.

A drawback is that, while changing the liquid level, the center of mass (CM) of the FM will move in the vertical direction. As we will see in the following section, this is not a major problem in our set-up since we never refer to the

CM for evaluating the gravitational force to compare with experimental data, but in all the cases we integrate over the volume of the liquid.

As an alternative to the liquid, we could also use a gas as FM. In this case the force should be modulated by changing gas pressure instead of liquid level. This would offer the big advantage that the CM of the FM does not change during modulation. The isobaric pressure gradient will also be reduced at negligible level. The greatest disadvantage is that, even at very high pressure, the gas density, and then the gravitational effect, are much smaller than for a liquid. For this reason, we adopt the liquid option, but in principle it is not excluded that in future a clever way can be found for using a modulated gas in gravity experiments.

The LAG principle of operation is described in detail in section 2.

In section 3, we discuss the application of this technique to the test of ISL and report results of numerical simulations showing how this can improve current upper limits on the $\alpha-\lambda$ exclusion plot [1][3] in the mm to cm region. We also underline the requirement for an accurate modelling of the apparatus to compare experimental data with predictions.

Conclusions are in section 4, where we also outline future steps and mention a starting R&D activity, funded by Italian National Institute of Nuclear Physics, to validate the principle of LAG experiment.

Finally, in the appendix we describe in detail how we compute the exclusion plot.

## 2. Principle of operation

The principle of operation of the LAG actuator is schematically shown in figure 1. Gravity force or torque acting on a TM is modulated by varying the liquid level in a tank placed (under vacuum) close to a TM suspended to a torsion pendulum. If a metallic tank is used, it will also serve as an electrostatic shield. The tank, which can be shaped as a cylinder or a parallelepiped, is connected through a pipe to an external liquid reservoir, where the liquid can be forced to flow in and out of the tank itself by means of a piston operated with a stepping motor or other kind of actuators (represented by a black box in figure). The variation of liquid level is performed at very low frequency (few mHz), so that it can be considered as almost stationary, and is controlled in such a way that the tank is never completely empty or full. In this way, only the liquid inside the tank is modulated while the liquid in the pipe does not contribute to modulate the gravity field. The only mechanically moving part in the experiment, that is the piston, can be placed several meters away from the apparatus, so that its gravitational effect at the modulation frequency is made completely negligible with respect to the effect of the liquid in the tank.

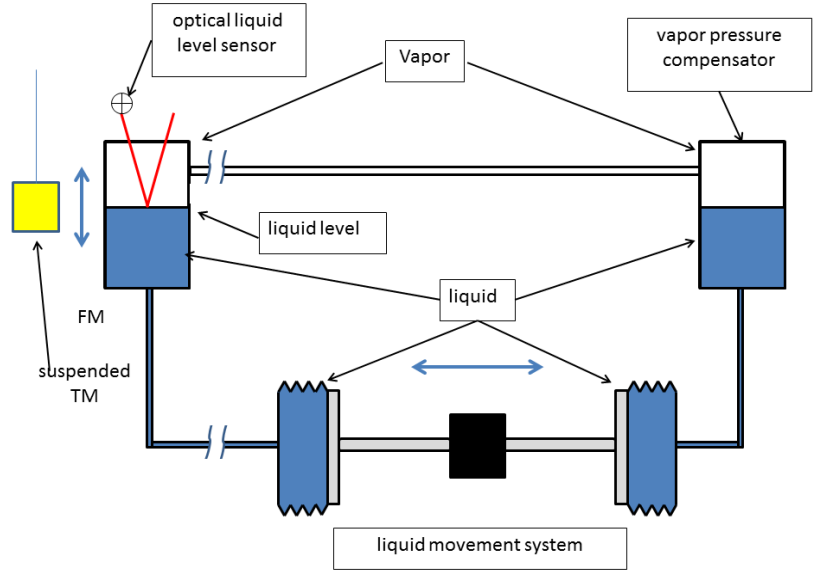

Figure 1 - Principle scheme of the LAG actuator

To avoid the presence and variation of other gasses diluted in the liquid when the liquid level is changed, the empty upper part of the tank must contain only vapour in equilibrium with its own liquid.

In order to keep the vapour pressure constant during operation, we will place a specular system, operated by the same piston in an opposite way, that forces liquid in a compensator tank (CT), as shown in figure 1. The upper part of the tanks are connected by a pipe, so that the vapour in both of them is at the same pressure. When we move the piston to change the liquid level in the FM, the level in the CT will vary in the opposite direction, so that the total volume (and pressure) of the gas remains (almost) constant.

Obviously, we cannot rely on the piston excursion to assess the liquid level in the tank. It must be accurately measured with a suitable sensor. We will use an optical lever sensor, as sketched in figure 1. A light beam is reflected at the liquid interface and the reflected beam is monitored by a position sensing device (PSD). With such a sensor, the liquid level can be measured with sub-micron accuracy [17] [18] and controlled in a repeatable way by acting in closed loop on the piston.

Such an actuator can be used in principle for any kind of gravitational experiment, like, for example, measurement of *G* or test of the equivalence principle. Size and set-up will obviously be adapted to any specific application. With suitable configurations (e.g. with a plane FM and a cylindrical TM), even short range applications, like sub mm ISL tests, are conceivable. For these tests, where electrostatic patch effects are known to be a disturbing factor, this scheme has an added benefit: the metallic liquid container acts as a natural electrostatic shield between TM and FM. Spurious electrostatic forces will still be present, but they represent a static background that has no effect at the modulation frequency.

In this paper, we analyse in detail the application of the LAG actuator to a test of ISL in the mm to cm distance range.

## 3. Inverse square law test

As a first application, we consider the test of ISL. In this type of experiments, gravity force (or torque) is measured at two (or more) FM-TM distances. Let's call $r_N$ and $r_F$ the near and far position and $F_N$ and $F_F$ the corresponding forces acting on the TM. The observable is the quantity $\gamma$ [3] defined (if we use the same point-like masses at the two positions) as:

$$\gamma = \frac{F_N}{F_F} \cdot \left(\frac{r_N}{r_F}\right)^2 \qquad (1)$$

where $\gamma = 1$ if Newtonian gravity holds; it is then convenient to represent deviations by introducing $\delta = 1 - \gamma$.
A deviation from Inverse Square Law (ISL) can be seen as a distance dependent gravity constant $G(r)$ [3] and the force between point-like masses is expressed as

$$F(r) = -G(r)\frac{Mm}{r^2} = -G_\infty \frac{Mm}{r^2}(1 + a(r)) \qquad (2)$$

This is usually parametrized by a Yukawa-like potential, a formalism introduced to represent an interaction mediated by a boson with a mass related to the interaction range $\lambda$.

$$V_{Yukawa}(r) = -G_\infty \frac{Mm}{r}\left(1 + \alpha e^{-r/\lambda}\right) \qquad (3)$$

No deviations from ISL have been experimentally observed so far, but upper limits have been set to the interaction strength $\alpha$ for a wide interval of the characteristic scale $\lambda$. It can be shown [3] that the experimental limit on $\alpha$ is set by the accuracy in the measurement of $\delta$, that in turn depends on both measurement errors and knowledge of the experimental apparatus parameters.

The present limits, for $\lambda$ above fraction of mm (a range accessible to the LAG actuator), are quite stringent, ranging from a few $10^{-3}$ to $2 \cdot 10^{-4}$. Assuming that the incertitude on all the other experimental parameters can be controlled below this accuracy level, the experimental S/N ratio in measuring $\delta$ should be at least of the same order of magnitude.

Another possible representation of deviation from ISL is the power law parametrization [3]. Arkani-Hamed, Dimopoulos, and Dvali (ADD) [7][9] introduced a model with *n* large extra dimensions ($n \geq 2$) to explain the so called hierarchy problem of the weakness of gravity with respect to other forces. It can be represented [3] by a potential of the form:

$$V_{power}(r) = -G_\infty \frac{Mm}{r}\left(1 + \left(\frac{\lambda}{r}\right)^n\right) \tag{4}$$

For *n* = 2, the Eotwash group has set a model independent upper limit to $\lambda < 44$ μm with a laboratory experiment [10]. The same data are reinterpreted in [3] setting a limit at 23 μm for the ADD model. For *n* > 2 (up to 6) current upper limits come from ATLAS data at LHC [11].

### 3.1 Proposed set-up

The set-up we propose is sketched in figure 2. We plan to use a two stage torsion pendulum like the apparatus, named PETER, that we developed for ground testing of the LISA-Pathfinder inertial sensor [20][21][22][23][24]. Such a system, with two soft degrees of freedom, allows simultaneous measurement of both the torque around the vertical axis and one horizontal component of the force acting on the TM and permits to constrain system geometry incertitude by measuring them at different relative TM-FM positions. In particular, with the set-up in figure 2, we can measure the component of the force along the y axis (the one parallel to the plate surface). In the LAG experiment, the cross shaped first suspension stage used in previous works will be replaced by a single bar hanging the TM at one end and a balancing mass at the other end. The TM will be a rectangular plate, while the balancing mass can be a sphere of the same mass, so that (ideally) no torque will be exerted on it by the gravity actuator.

An important aspect is that, while modulating force and torque, the TM will be controlled in feed-back by electrostatic actuation so that the distance and orientation with respect to the FMs will be kept constant during the measurement. The force and torque acting on the TM will then be extracted by the feed-back actuating signals.

It is worth noting that, while providing clear advantages in terms of wealth of signals and diagnostics, the use of a two stage pendulum is not strictly necessary, and a single stage pendulum, offering a simpler set-up and operation, could be used as well.

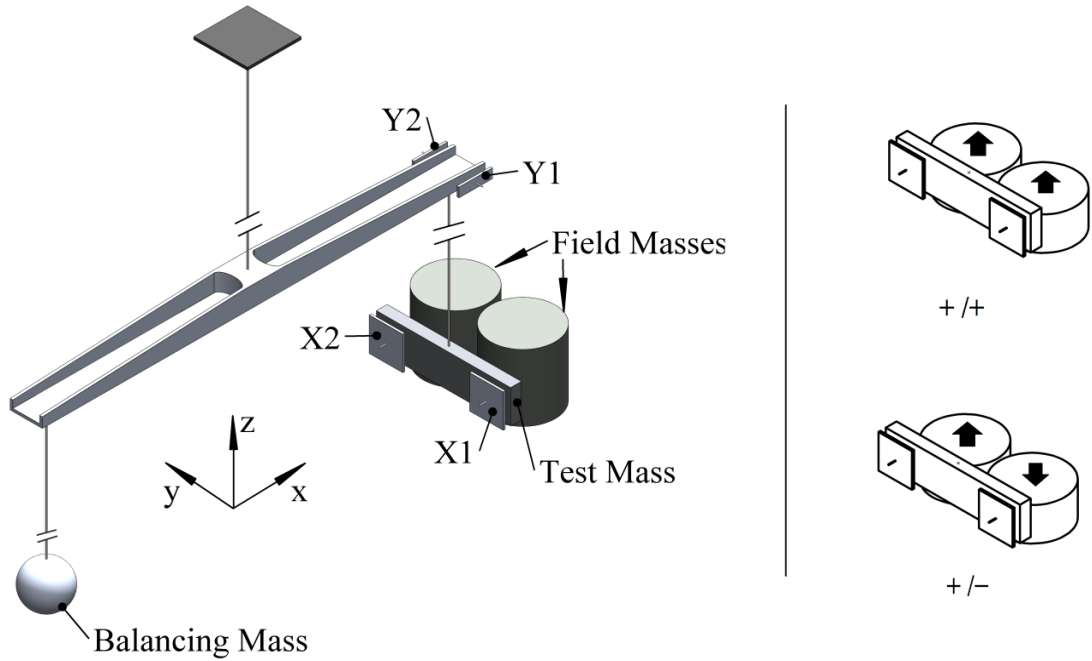

**Figure 2** - Sketch of the LAG experiment set-up (left). Schematic representation of the +/+ and +/- actuation modes (right). This configuration, with a two stage torsion pendulum, allows for simultaneous measurement of the torque around the vertical axis and one com component of the force acting on the TM. In this case, we actually measure the force along the y axis. X1 and X2 represent the electrodes used to control in feed-back the angular position of the TM around the vertical axis, while Y1 and Y2 are the ones used to control position along y (or equivalently the rotation of the upper bar around the vertical axis. Other TM-FM orientation with respect to the suspending bar can be adopted, permitting to measure a different component of the force, but we selected the most symmetric scheme in order to simplify the modelling of the effect of gravity on suspending bar and balancing mass.

The FMs will be composed by a pair of actuators as that described in section 2, carved from a single metallic block. The usage of two identical FMs that can independently actuate on the TM offer the advantage of several actuation modes:

1) The two FMs can be filled in phase, (i.e. at the same time), or in counter-phase (filling one while empting the other), this is represented by the +/+ or +/- sign in figure 2. This gives, for each relative TM-FMs position in x and y, different values of the force and torque, without changing the overall geometry.

2) Only one FM can be modulated, the other remaining stable (at any chosen level) and vice versa, in order to check the symmetry of the system.

3) The two FMs can be modulated at various amplitudes (liquid levels) and with various mean liquid heights.

From these measurements, we will be able to characterize possible asymmetries of the system (different FM radii, non-parallelism of cylinders, errors in cylinder axes orientation etc.).

The relative calibration of the two optical level sensors can be performed by connecting the two FMs in parallel (with a suitable valve manifold) and filling them together to the same liquid levels.

### 3.2 Evaluation of expected signals

Due to the tiny amplitude of the searched ISL deviations, an essential point is a reliable and precise evaluation of the expected Newtonian signal, based on the system geometry. This is not an easy task since the TM-FM relative distance is comparable with their size, so that they cannot be approximated with point-like masses. The total force/torque acting on the TM must then be computed by integrating the Newton force over the volume of all the involved masses. Due to the non-trivial shapes, there is no analytical solution, so the problem must be solved by numerical integration.

As a first step, in order to compute the order of magnitude of the expected signals in the proposed configuration, we started with a simplified model by dividing TM and FMs in elementary volumes and summing up to all pairs. In this way we neglect some details like the mechanical link of the TM to the suspending wire and the gravitational interaction with the other suspended masses of the pendulum (much smaller due to the larger distance, but not negligible), that should be taken into account for a more precise computation. Using this model, we can evaluate the expected force and torque as a function of relative TM-FMs position in the x-y plane and of the peak-to-peak liquid level amplitude. We assume pure mercury as a liquid and level amplitude modulation h = $\pm$ 25 mm at a frequency of 10 mHz. The TM is a molybdenum parallelepiped of length L= 100 mm, height 25 mm and thickness 8 mm, while the TMs radii are 24 mm, the distance between centers is D=L/2. A distance of 2 mm between TM and FMs surfaces was assumed in this computation.

Using this model, we can compute the expected force and torque as a function of relative TM-FMs position in the x-y plane and of the peak-to-peak liquid level amplitude. The results are shown in figure 3.

As we can see, for the +/+ actuation mode both the force along y and the torque are zero for y = 0. The force shows a maximum for y ~ $\pm$70 mm while the torque has a maximum at y ~ $\pm$ 55 mm. For the +/- mode both force and torque have the maximum in the centered position (y = 0 mm). These features will be very helpful to experimentally find the symmetry center of the TM-FMs system and to check its alignment and symmetry.

With the force and torque values shown in figure 3, the TM displacements at modulation frequency will be as large as ~ 50 μm and ~ 180 μrad; this confirms the need to operate the pendulum in closed-loop.

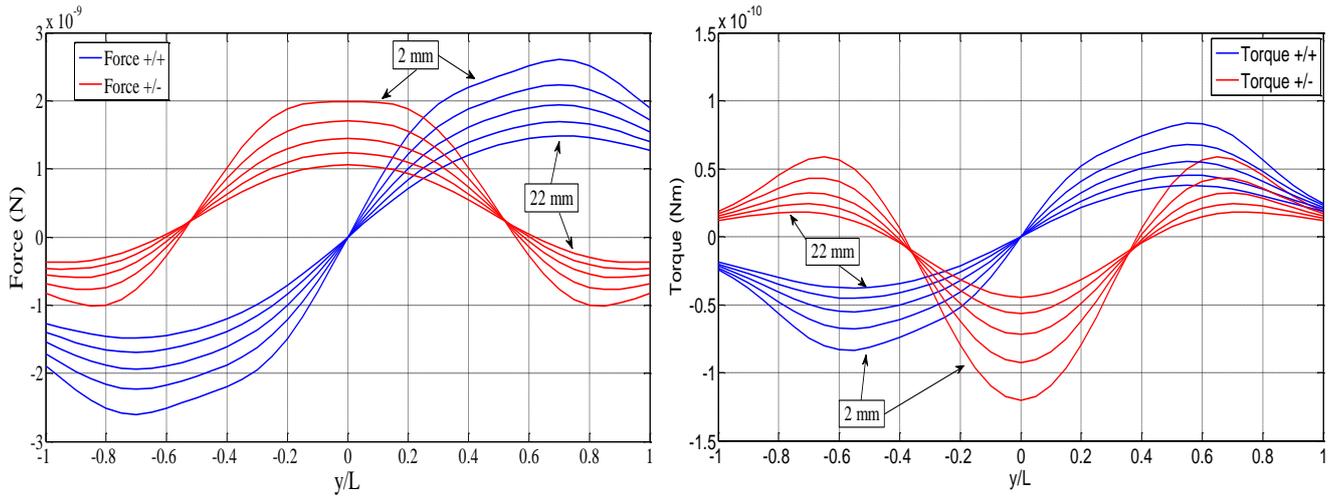

**Figure 3** - Force (left) and torque (right) amplitude versus lateral displacement (y) of the two FMs. For each configuration we compute the +/+ (blue) and the +/- (red) filling modes. In both cases, the 5 curve correspond to different TM-FM distance along x, starting from a minimum of 2 mm, with steps of 5 mm.

In figure 4 we report the force and torque power spectral density computed for the +/+ (blue) and +/- (red) actuation modes assuming sinusoidal liquid level variation at 10 mHz and an integration time of 80000 $s$, superimposed to the present sensitivity of PETER[22][24]. The amplitudes are computed, for two x distances (2 and 22 mm), at the y position that maximizes force and torque.

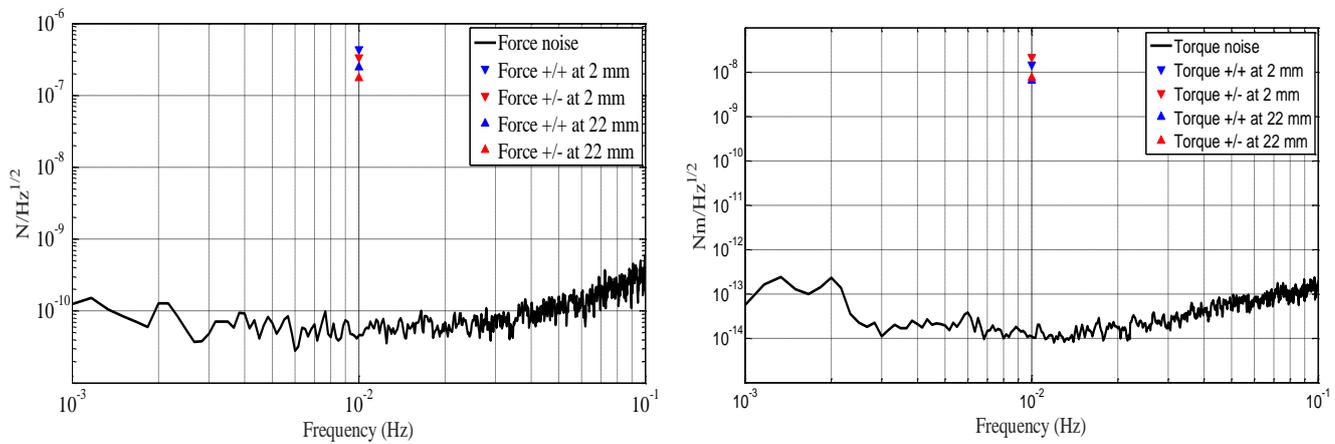

**Figure 4** - Power spectral density of force (left) and torque (right) produced by FM modulation described in the text, compared with the present sensitivity of the PETER facility. The expected S/N ratio is up to $10^4$ for force and above $10^5$ for torque.

As we can see, the expected S/N ratio exceeds $10^4$ for the force and $10^5$ for the torque. The amplitude of forces and torques produced by the LAG actuator can then be made large enough to perform state of art gravity experiments.

## 3.3 Expected upper limits on deviation from ISL

Figure 4 shows that the present torque sensitivity of $2 \cdot 10^{-14}$ N/Hz$^{1/2}$ allows a S/N ratio larger than $10^{-5}$ for the signal described above. This reflects in an uncertainty on $\delta$ of $10^{-5}$. In the following, we will assume the pendulum torque noise to be the limiting sensitivity factor in the described experiments. This is by no means a simple goal to achieve: a detailed error budget is needed, and it is currently being tackled, but is outside the scope of this paper. As examples of the possible sources of uncertainty that could affect the result, we cite: geometrical errors in positioning and moving the FMs, homogeneity, meniscus and pressure gradient of the liquid acting as FM, residual gas and thermal noise on the suspended TM, actuation noise in the TM feedback loop., Given this assumption, we can estimate the upper limits that this experiment can put to deviation to ISL. The computation is performed following the guideline reported in [3] for the case of extended bodies. A detailed description of the analysis is reported in appendix A

In figure 5, we compare, on the $\alpha-\lambda$ diagram, the current upper limits posed by previous experiments (sketched as magenta line) [10][12][13][14][15][16] with the limit that we aim to set with the LAG experiment in the proposed configuration. The limits are computed, according to equation (A12) in appendix A, for a set of measurements at distances ranging from 2 mm to 22 mm with steps of 5 mm. The assumed torque sensitivity (the one actually achieved with PETER [24]) is not limited by fundamental noises and can be potentially improved by more than one order of magnitude [26].

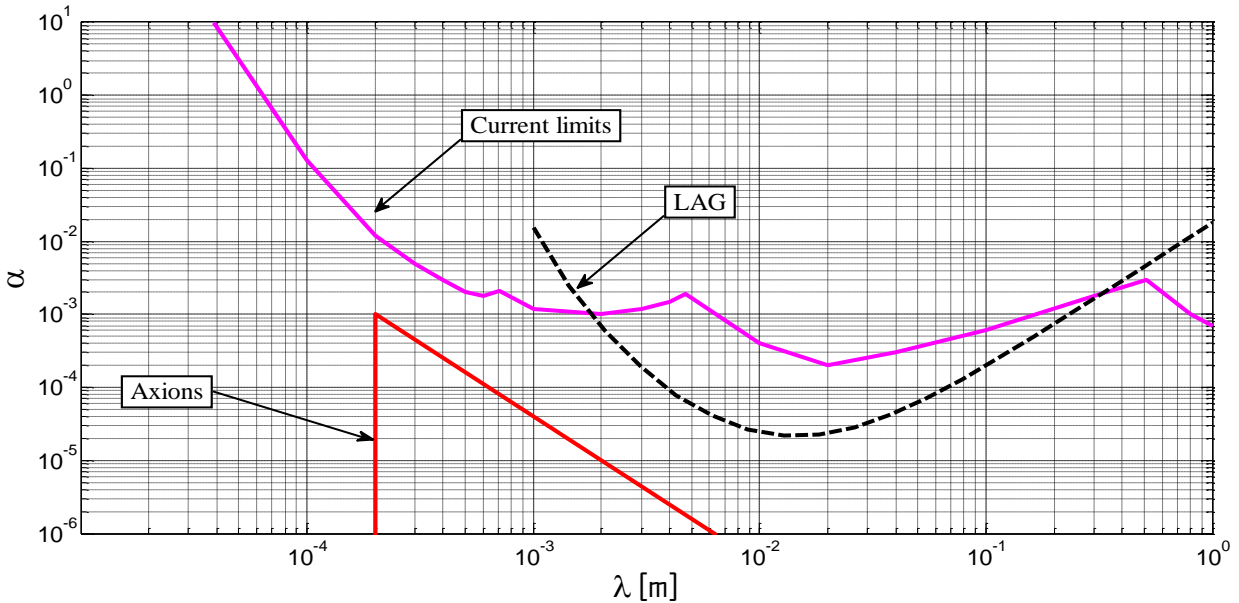

**Figure 5 Comparison of the current $\alpha-\lambda$ upper limits posed by previous experiments (magenta line) with the limit achievable by the LAG experiment (black dashed line) assuming, as a limiting factor, the present torque sensitivity of PETER. tTe limits are computed for a set of measurements ranging from a distance of 2 mm to a distance of 22 mm.**

As we can see, with the LAG experiment at the present torque sensitivity we can improve the exclusion plot by up to one order of magnitude in the $\lambda$ interval from few mm to tens of cm. In this range, possible deviations from ISL can be due to the existence of axions [1]; in figure 4 we also plot (red line) the corresponding allowed region (limited by non-gravitational experiments).

Concerning deviations from ISL described by a power law distance dependence, for the ADD model [17][18] with $n = 2$ we can compute the limit on $\lambda$ according to (A18) in appendix A, obtaining, with the present sensitivity, a new upper limits of $\lambda = 15$ μm with confidence level of 95 % ($2\sigma$).

## 4. Discussion and future steps

We have described a new actuation technique for laboratory gravity experiments, based on the modulation of the level of a liquid used as a FM, acting on a TM suspended to a torsion pendulum. By measuring the level of liquid with optical techniques and controlling in feed-back on the liquid movement system, (that is positioned far enough from the apparatus so that its effect on the TM is negligible. Force and/or torque on the TM can be modulated in a controlled and repeatable way, with a high S/N ratio.

We discussed a possible experiment to test the ISL that can potentially improve present limits on deviation from ISL law in the mm to cm region where interesting physics is represented by the possible existence of axions and by effects due to extra dimensions, like for the ADD model.

In principle, a similar device can be also used for other kind of gravity experiments, like, for example, an accurate measurement of the gravity constant $G$.

An R&D activity for the demonstration of the LAG principle of operation has been recently funded by Italian National Institute of Nuclear Physics (INFN). The prototype is currently being designed and will be tested with the PETER facility in the Napoli Gravity Physics Laboratory.

**Appendix**

A.1 - Introduction

In this appendix we describe in detail how we compute limits on deviation from ISL starting from the measured force and torque. In particular, we analyse the case of Yukawa and power low parametrizations introduced in section 3.

## A.2 – Yukawa parametrization

Deviation from ISL are usually parametrized by a Yukawa-like potential (eq. (3)), that for $\alpha = 0$ or $\lambda = \infty$, reduces to the Newtonian potential with $G$ not depending on distance. With this parametrization, the term $a(r)$ in eq. (2) can be written as:

$$a(r) = \alpha \left(1 + \frac{r}{\lambda}\right) e^{-r/\lambda} \tag{A1}$$

The force is then:

$$F(r, \alpha, \lambda) = -G_\infty \frac{Mm}{r^2}\left(1 + \alpha \left(1 + \frac{r}{\lambda}\right) e^{-r/\lambda}\right) \tag{A2}$$

we can compute the exclusion plot by substituting (A1) into (1), and solving for $\alpha$:

$$\alpha = \frac{\delta e^{r_N/\lambda}}{\left(1 + \frac{r_N}{\lambda}\right) - (\delta + 1)\left(1 + \frac{r_F}{\lambda}\right) e^{-(r_F - r_N)/\lambda}} \tag{A3}$$

here $\delta$ is inserted with its error. A measured value of $\delta$ different from 0 (within experimental accuracy) implies deviation from ISL. If it is compatible with zero, the accuracy on $\delta$ gives provides for the value of $\alpha$ at the minimum in $\lambda$ [3].

### A.2.1 – Finite size effects

In the case of extended TM and FM, equation (A2) must be replaced by:

$$F(r_i, \alpha, \lambda) = -G_\infty \rho_{TM} \rho_{FM} \int_{V_{TM}} dV_{TM} \int_{V_{FM}} dV_{FM} \frac{1}{r_i^2}\left(1 + \alpha \left(1 + \frac{r_i}{\lambda}\right) e^{-r_i/\lambda}\right) \tag{A4}$$

Where the index $i$ can take the values $i = N, F$.

Reference [1] reports the case of infinite plates, where the integrals (6) can be solved analytically. In the more general case, we must solve these integrals numerically (so a good knowledge of the geometry of the system is essential).

It is useful to define:

$$A_i = \int_{V_{TM}} dV_{TM} \int_{V_{FM}} dV_{FM} \frac{1}{r_i^2}\left(1 + \frac{r_i}{\lambda}\right) e^{-r_i/\lambda} \tag{A5}$$

and

$$R_i = \int_{V_{TM}} dV_{TM} \int_{V_{FM}} dV_{FM} \frac{1}{r_i^2} \tag{A6}$$

So that we can get:

$$F_N = F(r_N, \alpha, \lambda) = -G_\infty \rho_{TM} \rho_{FM} (R_N + \alpha\, A_N), \quad F_F = F(r_F, \alpha, \lambda) = -G_\infty \rho_{TM} \rho_{FM} (R_F + \alpha\, A_F) \tag{A7}$$

And

$$F_N^0 = F(r_N, \alpha = 0, \lambda = \infty) = -G_\infty \rho_{TM} \rho_{FM} (R_N), \quad F_F^0 = F(r_F, \alpha = 0, \lambda = \infty) = -G_\infty \rho_{TM} \rho_{FM} (R_F) \tag{A8}$$

Where the $A_i$ terms in (7) give the spatial dependence of the force, while the terms $R_i$ in (8) give the same in the case of Newtonian gravity.

We can now define $\gamma$ (or $\delta = \gamma - 1$) as:

$$\gamma = \frac{G_N}{G_F} = \frac{F_N}{F_F} \cdot \frac{F_F^0}{F_N^0} = \frac{(R_N + \alpha\, A_N)}{(R_F + \alpha\, A_F)} \cdot \frac{R_F}{R_N} = \frac{1 + \alpha \frac{A_N}{R_N}}{1 + \alpha \frac{A_F}{R_F}} \tag{A9}$$

It is worth noting that the ratio $R_F/R_N$ in (A9) plays the same role as the ratio $(r_F/r_N)^2$ in (2), representing the ISL slope of the Newtonian gravity.

As for the point-like case, we can solve equation (A9) for $\alpha$:

$$\alpha = \frac{\delta}{\frac{A_N}{R_N} - (\delta + 1)\left(\frac{A_F}{R_F}\right)} \tag{A10}$$

Where $\delta$ is the result of the measurement while $R_N$ and $R_F$ must be computed numerically for each position and $A_N$ and $A_F$ must be computed for each position and for each value or $\lambda$.

As expected, (A10) reduces to (A3) in the case of point-like masses.

A.3 – Power law parametrization

In the case of power law parametrization, $a(r)$ in eq. (1) becomes:

$$a(r) = (1 + n)\left(\frac{\lambda}{r}\right)^n \tag{A11}$$

And the force becomes:

$$F(r, n, \lambda) = -G_\infty \frac{Mm}{r^2}\left(1 + (1 + n)\left(\frac{\lambda}{r}\right)^n\right) \tag{A12}$$

For the point-like case, substituting (A12) in (1), we obtain:

$$\lambda = \left(\frac{\delta/(1+n)}{\left(\frac{1}{r_N}\right)^n - (\delta + 1)\left(\frac{1}{r_F}\right)^n}\right)^{1/n} \tag{A13}$$

This allows, for a given experimental value of $\delta$, to constrain the range $\lambda$ for a model with any value of n (assumed as an integer representing the number of extra dimensions in the ADD model).

In the case of extended TM and FM (A12) becomes:

$$F(r_i, n, \lambda) = -G_\infty \rho_{TM} \rho_{FM} \int_{V_{TM}} dV_{TM} \int_{V_{FM}} dV_{FM} \frac{1}{r_i^2} \left(1 + (1+n)\left(\frac{\lambda}{r}\right)^n\right) \quad (A14)$$

In analogy with the Yukawa parametrization, we can define:

$$B_i = \int_{V_{TM}} dV_{TM} \int_{V_{FM}} dV_{FM} \frac{1}{r_i^{2+n}} \quad (A15)$$

And, using the previously defined $R_i$ (eq. (A6)), we get:

$$F_N = F(r_N, n, \lambda) = -G_\infty \rho_{TM} \rho_{FM}(R_N + (1+n)\lambda^n B_N), \quad F_F = F(r_F, \alpha, \lambda) = -G_\infty \rho_{TM} \rho_{FM}(R_F + (1+n)\lambda^n B_F) \quad (A16)$$

And

$$F_N^0 = F(r_N, \lambda = 0) = -G_\infty \rho_{TM} \rho_{FM}(R_N), \qquad F_F^0 = F(r_F, \lambda = 0) = -G_\infty \rho_{TM} \rho_{FM}(R_F) \quad (A17)$$

And as in (A13) for the point like case, we can solve for $\lambda$:

$$\lambda = \left(\frac{\delta/(1+n)}{\frac{B_N}{R_N} - (\delta+1)\frac{B_F}{R_F}}\right)^{1/n} \quad (A18)$$

A.4 – Advantage of a two DOFs system

In the case of a two DOFs system, we can make simultaneous measurements of force ($F$) and torque ($\tau$) acting on the TM. Both force and torque measurement can provide independent upper limits on $\alpha$; let's call them $\alpha_F$ and $\alpha_\tau$.

To compute $\alpha_\tau$ we can use the same approach as in (6) but integrating on the torque that the FM exerts on the TM. We can define:

$$\gamma_\tau = \frac{G_N}{G_F} = \frac{\tau_N}{\tau_F} \cdot \frac{\tau_F^0}{\tau_N^0} = \frac{(R_{\tau N} + \alpha_\tau A_{\tau N})}{(R_{\tau F} + \alpha_\tau A_{\tau F})} \cdot \frac{R_{\tau F}}{R_{\tau N}} = \frac{1 + \alpha_\tau \frac{A_{\tau N}}{R_{\tau N}}}{1 + \alpha_\tau \frac{A_{\tau F}}{R_{\tau F}}} \quad (A19)$$

and derive an expression similar to the eq. (A10):

$$\alpha_\tau = \frac{\delta_\tau}{\frac{A_{\tau N}}{R_{\tau N}} - (\delta_\tau + 1)\left(\frac{A_{\tau F}}{R_{\tau F}}\right)} \quad (A20)$$

The volume integrals $A_{\tau i}$ and $R_{\tau i}$ have different values than those computed for the force. For each considered experimental configuration (relative TM-FM orientation) and each value of lambda, it is worth exploring whether the force or the torque measurement provides the best limit on $\alpha$.

**Acknowledgements**

This work is supported by INFN scientific commission V.